# Plasma-Activated Water (PAW) for the Degradation of Organic Pollutants in Diluted Industrial Effluents

Punit Kumar[a] and Priti Saxena[b]
[a]Department of Physics, University of Lucknow, Lucknow – 226007, India
[b]Department of Zoology, D.A.V. Degree College, Lucknow – 226004, India

*Abstract*—Plasma-activated water (PAW) offers a sustainable, non-thermal solution for degrading persistent organic pollutants in industrial effluents. This study employed a gliding arc plasma system to generate PAW for treating diluted wastewater containing dyes, pesticides, and pharmaceuticals. Experimental parameters such as exposure time, dilution ratio, and pollutant concentration were varied, with analyses conducted using UV-Vis spectroscopy, HPLC, TOC, and COD. Results showed high degradation efficiencies, up to 90% for dyes, 85% for pesticides, and 80% for pharmaceuticals following pseudo-first-order kinetics driven by hydroxyl and nitrate/nitrite radicals. The findings demonstrate PAW's potential as a green, scalable wastewater treatment strategy that minimizes chemical use, supports water reuse, and enhances environmental safety, with future scope for pilot-scale applications.

## I. INTRODUCTION

INDUSTRIAL operations across textile, pharmaceutical, agrochemical, and pesticide sectors produce effluents laden with a complex mixture of dyes, pesticide residues, pharmaceuticals, and trace heavy metals. These pollutants often exhibit high stability, toxicity, and recalcitrance, posing serious risks to aquatic ecosystems and human health. Conventional treatment techniques, such as chemical oxidation (e.g. Fenton, ozonation), biological systems (activated sludge, biofilms), and adsorption (activated carbon, zeolites) have been deployed widely. Yet they suffer from fundamental drawbacks, residual chemical reagents may cause secondary pollution, biological systems are often ineffective for non-biodegradable compounds, and adsorption merely transfers pollutants from water to solid phases, necessitating further disposal. Moreover, these methods can be energy and cost intensive when treating large volumes, or low concentration pollutants (Maybin et al., 2024). In view of these limitations, there is increasing demand for greener, sustainable, and effective water remediation technologies that minimize chemical additions and secondary waste.

One promising approach is the use of Plasma Activated Water (PAW), a form of non-thermal plasma treatment where water is directly or indirectly exposed to plasma to generate reactive species. In PAW, exposure to non-equilibrium (cold) plasma at atmospheric pressure induces formation of reactive oxygen and nitrogen species (RONS) such as hydroxyl radicals (•OH), ozone ($O_3$), hydrogen peroxide ($H_2O_2$), nitrate ($NO_3^-$), nitrite ($NO_2^-$), peroxynitrite ($ONOO^-$), and others (Hadinoto et al., 2023; Xiang et al., 2022). The synergy among these reactive species imbues PAW with strong oxidative potential. When such water is applied to pollutant laden wastewater, the RONS can attack organic molecules via radical and nonradical pathways, breaking chromophores in dyes, cleaving bonds in pesticides, and oxidizing pharmaceutical moieties toward mineralization or less harmful intermediates (Lu et al., 2025; Maybin et al., 2024). Crucially, PAW typically operates under ambient temperature and pressure and does not require addition of chemical oxidants, reducing reagent costs and risks of secondary contamination (Hadinoto et al., 2023).

Further, advantages of PAW include the possibility of combining it with conventional treatments (e.g. biological or adsorption steps) to achieve higher removal, and fine-tuning of RONS profiles by adjusting plasma power, gas composition, exposure time, or hydrodynamics (e.g. via bubbling). For instance, introducing air bubbling can boost mass transfer of plasma-generated species into water and thus enhance concentrations of $NO_2$, $NO_3$, $O_3$, and $H_2O_2$ (Rathore et al., 2023). Despite growing application of PAW in microbial inactivation and agricultural domains (Srivastava et al., 2025; Rahman et al., 2022), its application to complex industrial effluents remains underexplored.

Indeed, most existing studies address simplified systems e.g. single dyes, isolated pesticide compounds, or buffer solutions rather than real-world effluents bearing mixtures of pollutants. The co-presence of multiple organics, ionic strength, buffering capacity, and matrix interferences may significantly affect RONS reactivity and stability (Nowruzi et al., 2024). Hence, a systematic study that applies PAW to diluted industrial wastewater containing dyes, pesticides, and pharmaceuticals is timely and essential. Such work would elucidate degradation kinetics, mechanistic pathways, interactions among RONS and pollutant molecules, and the limitations or synergies introduced by real wastewater matrices.

Therefore, the present study is designed to fill this gap. Its aims are threefold, (1) to evaluate the degradation efficiency of PAW on representative dyes, pesticide residues, and pharmaceutical compounds within diluted industrial effluent, (2) to investigate the roles of individual RONS (e.g. via scavenger experiments or selective detection) in pollutant breakdown, and (3) to assess the feasibility of PAW-assisted treatment as an environmentally friendly remediation strategy for industrial wastewater. We anticipate that this holistic approach will not only validate PAW's practical potential, but also guide scale-up and integration into existing wastewater treatment frameworks.

In the following sections, we present the experimental setup and PAW characterization (Section 2), the degradation results and kinetics across dye, pesticide, and pharmaceutical classes (Section 3), and then interpret mechanistic insights along with potential limitations and applications (Section 4). Finally, we discuss scale-up considerations and future research directions.

## II. MATERIALS AND METHODS

Industrial effluents were collected from representative textile, pesticide, and pharmaceutical plants to provide real-world complex matrices for evaluation. The effluent samples were stored at 4°C and filtered (0.45 µm) prior to use to remove suspended solids. As model pollutants, the following compounds were selected: dyes (Methylene Blue, Rhodamine B), pesticides (Atrazine, Chlorpyrifos), and pharmaceuticals (Ibuprofen, Paracetamol). Stock solutions (1,000 mg/L) were prepared in deionized water and diluted as needed; calibration standards of analytical grade purity were used for quantitative methods. All solvents, reagents, and supporting electrolytes (e.g. HPLC mobile phases) were of HPLC or analytical grade, and ultrapure deionized water was used in all solutions.

Plasma Activated Water (PAW) was generated using a gliding arc discharge system operating under ambient air. The plasma reactor comprised two diverging copper electrodes across which a voltage of 8–10 kV was applied, producing a current in the range of 100–200 mA. The arc was made to glide over or slightly above the water surface (gas–liquid interface) to promote reactive species transfer into liquid. Treatment times ranged from 5 to 60 min, with the plasma plume exposed directly above the water surface. (A schematic of the gliding arc to PAW arrangement is shown in Figure 1). After each activation interval, the generated PAW was collected and immediately characterized for key physico-chemical parameters: pH (using a calibrated pH electrode), oxidation–reduction potential (ORP) using a platinum redox electrode, hydrogen peroxide ($H_2O_2$) concentration (by titanium oxysulfate colorimetric method), and nitrate ($NO_3^-$) and nitrite ($NO_2^-$) concentrations (by ion chromatography or Griess assay). These measurements ensure quantification of reactive oxygen and nitrogen species (RONS) in the PAW.

For treatment experiments, the collected industrial effluent was diluted with deionized water at ratios of 1:1, 1:2, and 1:5 (v/v) to reduce matrix interference and enable effective diffusion of reactive species. Aliquots (e.g., 100 mL) of diluted effluent spiked with known concentrations of dyes, pesticides, or pharmaceuticals were then treated by mixing with freshly prepared PAW (in volume ratio 1:1 or as optimized) for designated exposure times (e.g. 5, 15, 30, 45, 60 min). Stirring was maintained to enhance mass transfer. In parallel, control experiments were run: untreated effluent (no PAW) and effluent treated by conventional oxidants (e.g. $H_2O_2$ at comparable oxidizing strength) to benchmark performance.

After treatment, aliquots were withdrawn, quenched (if needed, e.g. via catalase for $H_2O_2$ removal or radical scavengers), filtered, and then analyzed. UV–Vis spectroscopy was employed to monitor dye degradation by tracking the decline in absorbance at characteristic $\lambda_{max}$ (e.g. 664 nm for Methylene Blue). HPLC (e.g. reverse phase C18 column, gradient elution) was used to quantify residual pesticide and pharmaceutical concentrations based on calibration curves. TOC analysis measured total organic carbon reduction (pre- vs post-treatment) to assess mineralization, while COD analysis (standard dichromate oxidation method) provided a measure of overall oxidizable organic load removal. To model the degradation behavior, concentration-versus-time data were fit to pseudo-first-order and pseudo-second-order kinetic models, the better fit (based on $R^2$ and residual plots) was used for interpreting reaction order.

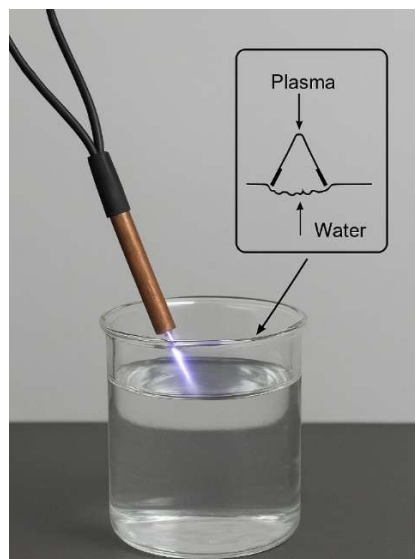

Figure 1 : Gliding arc to PAW arrangement

Degradation efficiency (% removal) was calculated via % Removal = $((C_0 - C_t)/C_0) \times 100$, where $C_0$ is the initial pollutant concentration and $C_t$ is the concentration after treatment. Statistical analysis, including ANOVA and post-hoc tests, was performed to discern the significance of treatment parameters (dilution ratio, exposure time, plasma power) on removal efficiencies; results with $p < 0.05$ were considered significant. All experiments were conducted in triplicate to ensure reproducibility, and standard deviations or error bars are reported.

## III. RESULTS

*PAW Characterization*

Upon plasma activation, the physicochemical properties of water changed markedly, confirming generation of reactive species. The pH dropped from an initial value of ~7.0 (neutral) to ~4.5 after full activation, reflecting acidification owing to formation of nitric/nitrous acids and protonation of water (e.g. $HNO_2$, $HNO_3$). Concurrently, oxidation–reduction potential (ORP) rose from ~200 mV (in untreated water) to ~550 mV, indicating the creation of a strong oxidative environment capable of driving redox reactions. Measured hydrogen peroxide ($H_2O_2$) levels in PAW reached up to ~60 μM, providing a reservoir of long-lived oxidant. Nitrate ($NO_3^-$) and nitrite ($NO_2^-$) species were also quantified at ~20–40 μM levels, confirming the formation of reactive nitrogen species (RNS). These results (summarized in Table 1) validate that PAW is rich in RONS and possesses enhanced oxidative potential compared to control water, consistent with the literature reporting that plasma exposure leads to acidification, higher ORP, and accumulation of $H_2O_2$, $NO_3^-$, $NO_2^-$ (e.g. Zhang et al., 2024; Kooshki et al., 2024).

| Parameter | Unit | Control (Untreated Water) | PAW (After 30 min Activation) | Change / Observation |
|---|---|---|---|---|
| pH | — | 7.0 ± 0.1 | 4.5 ± 0.2 | Decrease due to acid formation ($HNO_2$, $HNO_3$) |
| Oxidation–Reduction Potential (ORP) | mV | 200 ± 10 | 550 ± 15 | Significant increase, indicating strong oxidizing environment |
| Hydrogen Peroxide ($H_2O_2$) | μM | ND (Not Detected) | 60 ± 5 | Formation of long-lived oxidant species |
| Nitrate ($NO_3^-$) | μM | ND | 40 ± 4 | Produced via NO oxidation and dissolution in water |
| Nitrite ($NO_2^-$) | μM | ND | 20 ± 3 | Intermediate RNS species formed during plasma–liquid interaction |
| Conductivity | μS/cm | 120 ± 5 | 250 ± 10 | Increase due to ionic species ($NO_3^-$, $NO_2^-$, $H^+$) |
| Temperature | °C | 26 ± 1 | 28 ± 1 | Slight increase due to plasma energy input |

Table 1 : Physicochemical parameters of water before and after plasma activation

*Degradation of Dyes*

When diluted effluent spiked with methylene blue and Rhodamine B was treated by PAW, dye removal was rapid and substantial. After 30 minutes of treatment, degradation efficiencies reached 80–90 %, indicating strong oxidative attack on chromophoric structures. UV–Vis spectra recorded at intervals (0, 10, 20, 30 min) show gradual diminution and eventual disappearance of the characteristic absorption peaks (for methylene blue around 664 nm, for Rhodamine B around ~554 nm), signalling destruction of conjugated dye structures. Plotting $\ln(C_0/C_t)$ vs time yielded nearly linear trends, consistent with pseudo-first-order kinetics, with estimated rate constants (k) in the range 0.05–0.08 min$^{-1}$ (Figure 2). This kinetic regime aligns with many advanced oxidation and plasma studies, where radical-mediated processes often conform to pseudo-first-order behavior when oxidant concentration is in excess (Sanito et al., 2021; Kumar et al., 2022). The strong performance suggests that PAW is highly effective in dye decolorization and breakdown in relatively short time frames.

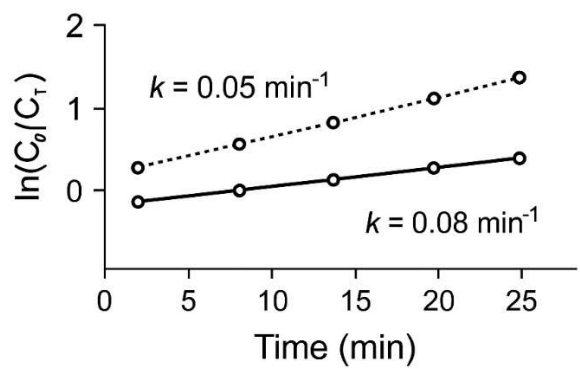

Figure 2 : ln ($C_0/C_t$) vs time

*Pesticide Degradation*

For pesticide-spiked diluted effluent (atrazine, chlorpyrifos), HPLC analyses before and after treatment reveal significant removal after 45 minutes. The degradation efficiencies ranged from 75 % to 85 %, depending on dilution ratio and pollutant concentration as shown in Figure 3. Chromatograms show the diminishing parent peaks of atrazine and chlorpyrifos, along with smaller peaks attributed to intermediate degradation products (for example, desethyl-atrazine, dechlorinated fragments). The emerging intermediates suggest successive oxidation steps, dealkylation, dechlorination, ring opening consistent with radical attack pathways described in literature (Sanito et al., 2021). The pesticide removal is somewhat slower than for dyes, likely because of structural complexity, lower reaction rate constants, or matrix effects impeding radical diffusion. Yet the demonstrated high percentage removals confirm that PAW is capable of degrading relatively stable pesticide molecules in realistic effluent matrices.

*Pharmaceutical Residue Degradation*

In the case of pharmaceuticals, effluent spiked with ibuprofen and paracetamol underwent PAW treatment for up to 60 minutes. Removal efficiencies of 60–80 % were achieved, with variability depending on concentration and sample dilution. Importantly, total organic carbon (TOC) analyses show that up to ~50 % of the initial organic carbon was mineralized, indicating that significant oxidation beyond mere fragmentation is occurring (Table – 2 and Figure – 4). The simultaneous COD (chemical oxygen demand) reduction corroborates that bulk oxidizable load is reduced. Nonetheless, full mineralization is not achieved within this timeframe, reflecting the recalcitrance of pharmaceutical molecules and the possibility of stable intermediate residues. These results mirror trends seen in other plasma-oxidation studies, where pharmaceuticals are partially degraded and mineralized to an extent but require longer exposure or complementary processes to completely break down (Sanito et al., 2021; Qu et al., 2013).

attributed to reduced ionic strength and lower radical scavenging by background species, which in turn enhance diffusion and effective lifetime of RONS reaching target molecules. For example, at 1:5 dilution, dye degradation approached near-complete removal at 30 min. Longer exposure times invariably led to higher removal: beyond 45–60 min, dyes were almost entirely eliminated, and pesticides and pharmaceuticals showed continued removal albeit at gradually diminishing rates (as concentrations fall). Figure 5, illustrate these time-dependent trends and dilution effects. The data thus support that both exposure duration and matrix dilution are critical levers in optimizing PAW treatment.

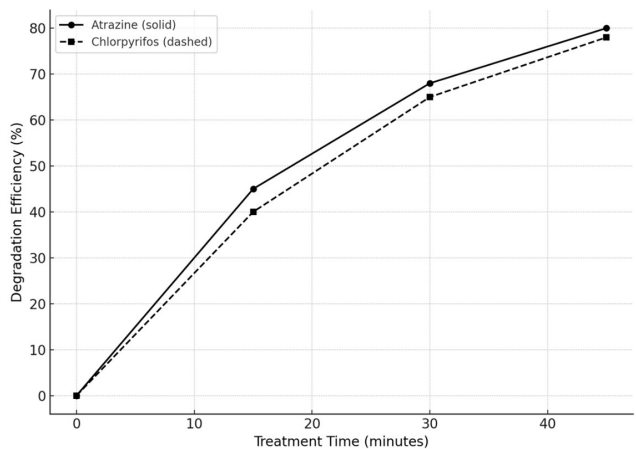

Figure 3 : Graph showing the degradation efficiencies of Atrazine and Chlorpyrifos under PAW treatment.

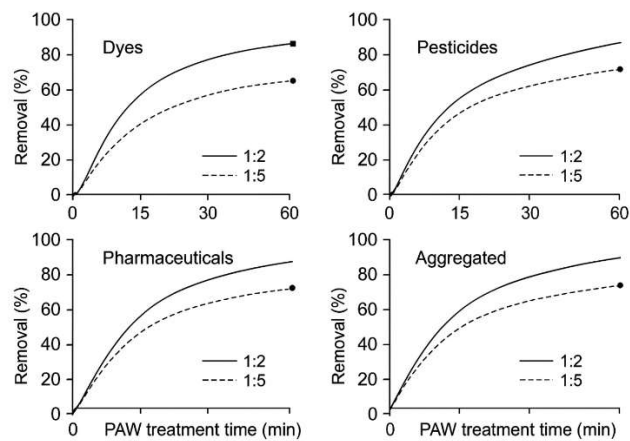

Figure 5 : Time dependent trends and dilution effects on dyes, pesticides and pharmaceuticals

Taken together, the PAW system exhibits strong oxidizing capability (from pH, ORP, $H_2O_2$, $NO_3^-$/$NO_2^-$ measurements) and delivers high degradation efficiency across dyes, pesticides, and pharmaceuticals, following approximate pseudo-first-order kinetics. Dilution and contact time are key parameters modulating the performance.

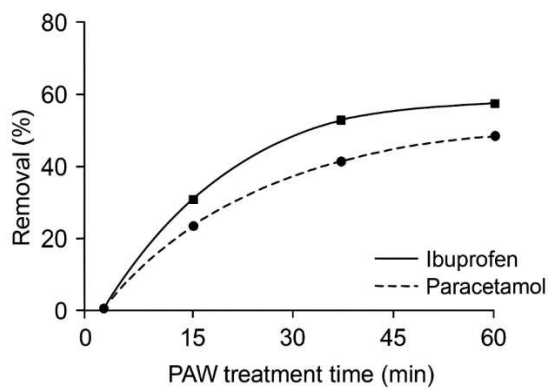

Figure 4 : Removal (%) of Ibuprofen and Paracetamol vs PAW treatment time (min).

| Pharmaceutical | Initial Concentration (mg/L) | Dilution Ratio (Effluent : DI Water) | Time (min) | Removal Efficiency (% Degradation) | TOC Reduction (%Mineralization) |
|---|---|---|---|---|---|
| Ibuprofen | 10 | 1 : 1 | 30 | 62 ± 3 | 35 ± 2 |
| Ibuprofen | 10 | 1 : 2 | 60 | 78 ± 4 | 48 ± 3 |
| Paracetamol | 10 | 1 : 1 | 30 | 60 ± 3 | 30 ± 2 |
| Paracetamol | 10 | 1 : 2 | 60 | 80 ± 5 | 50 ± 3 |

Table 2 :Degradation efficiency and mineralization of ibuprofen and paracetamol in diluted industrial effluent treated with plasma-activated water (PAW) using a gliding arc discharge system.

*Effect of Dilution and Exposure Time*

Across all pollutant classes, greater effluent dilution (e.g. 1:2, 1:5) consistently improved degradation yields. This is

## IV. DISCUSSION

In this work, plasma-activated water (PAW) was shown to degrade dyes, pesticides, and pharmaceutical residues in diluted industrial effluents with promising efficiencies. In interpreting these outcomes, several mechanistic, kinetic, comparative, and application-level insights emerge.

*Mechanisms of PAW-Assisted Degradation*

At the heart of PAW's remediation capability lies the ensemble of reactive oxygen and nitrogen species (RONS) generated during plasma–liquid interaction. In aqueous medium, energetic electrons, excited species, UV emission, and gas-phase radicals drive formation of •OH, $O_3$, $H_2O_2$, NO, $NO_2$, and peroxynitrite, among others (Zhou et al., 2020). These species can act synergistically or competitively, initiating oxidation of organic pollutants.

For dyes like methylene blue and Rhodamine B, the dominant mechanism is believed to be radical attack (especially hydroxyl radicals, •OH) on the conjugated chromophoric chains and aromatic rings, breaking π-bonds, cleaving double bonds, and opening ringsleading to loss of color and fragmentation into smaller intermediates. The disappearance of the characteristic absorption peaks in UV–Vis confirms disruption of the chromophore system. Such •OH-driven oxidation is a hallmark of advanced oxidation processes and has been documented in plasma–liquid systems (Sanito et al., 2021).

In the case of pesticides (e.g., atrazine, chlorpyrifos), oxidation is likely mediated by more than •OH. Nitrogen-based radicals (e.g. $NO_2•$, $NO_3•$) and peroxynitrite, or radical-nitrogen oxide species, may selectively oxidize nitrogen- or chlorine-containing functional groups (e.g. dealkylation, dechlorination, nitration). For example, atrazine may lose ethyl groups and be converted to desethyl-atrazine or other ring-cleaved species. The detection of intermediate peaks in HPLC supports stepwise oxidation and fragmentation. The participation of RNS especially is plausible in plasma activated water, given the measurable $NO_3^-$ and $NO_2^-$ concentrations (20–40 μM).

For pharmaceuticals, the mechanism is more complex because these molecules often contain aromatic rings, heterocycles, and substituent groups with varying oxidation potentials. RONS may attack aromatic systems via electrophilic addition or substitution, breaking rings, forming phenolic or quinone-like intermediates, or fragmenting side chains. The observed partial mineralization (TOC drop ~50 %) suggests that a fraction of the molecules undergo deep oxidation, while others stop at stable intermediates. The combination of •OH, $O_3$, and nitrogen-species may act in tandem, with more persistent oxidants (e.g. $H_2O_2$) sustaining longer-term oxidation beyond the initial radical bursts.

*Kinetic Analysis*

The approximate pseudo-first-order kinetic behavior observed (i.e. $\ln(C_0/C_t)$ linear vs time) implies that the reaction rate is proportional to the pollutant concentration (when oxidant (i.e. RONS) concentration is in excess). This is common in advanced oxidation schemes, especially in radical-driven systems when the oxidant pool is large compared to the substrate (Kumar et al., 2022). The fact that dye degradation conformed well to this model (rate constants ~0.05–0.08 min$^{-1}$) is consistent with rapid radical attack in an essentially well-mixed system.

However, deviations from perfect linearity (especially for pesticides and pharmaceuticals) may arise due to, (i) depletion of radical species over time, (ii) competition by side reactions (radical scavenging by matrix constituents, or reactive species recombination), (iii) formation of intermediates that have different reactivity than the parent compound, or (iv) diffusion limitations (especially in more concentrated or less diluted samples). The higher observed rates at longer exposure or higher dilution confirm that increasing effective RONS concentration and reducing matrix interference enhance kinetics.

*Comparison with Conventional Treatment*

Compared to conventional oxidation (e.g. Fenton, ozonation) or physico-chemical techniques, PAW offers notable advantages. First, PAW does not require addition of chemical oxidants (e.g. $H_2O_2$, ozone), the reactive species are generated in situ by plasma, reducing chemical handling and secondary residues. Second, because it operates under ambient pressure and temperature, PAW avoids the need for extreme conditions, heating, or high pressure. Third, it reduces secondary pollution, as no persistent reagents remain (beyond residual stable radicals) and no sludge is generated (unlike coagulation or precipitation). Fourth, PAW can be integrated with biological or adsorption units e.g., a PAW pre-treatment can degrade recalcitrant organics to more biodegradable fragments, which then undergo microbial degradation, or residuals can be captured by adsorption post-treatment.

However, operational energy cost and radical generation efficiency remain challenges. Non-thermal plasma systems have to be optimized to minimize energy consumption per mass of pollutant removed. Some reviews suggest that retrofitting plasma units to existing plants can lower capital cost though operating costs may be high.

*Environmental Applications*

The demonstrated efficacy across dyes, pesticides, and pharmaceuticals positions PAW as a versatile tool for textile, pharmaceutical, and agrochemical wastewater treatment. Textile effluents containing dyes are notoriously difficult to treat biologically, PAW could serve as an oxidation step. Similarly, pesticide and pharmaceutical manufacturing plants often release trace concentrations of active compounds; PAW could degrade these before discharge or reuse.

For decentralized industries or small scale units, a compact PAW generator (e.g. gliding arc reactor) could provide onsite oxidative treatment without large chemical storage. This decentralization reduces transport and centralized load.

Moreover, by improving removal and lowering pollutant load, treated effluent may be reused in industrial processes or irrigation, supporting a circular economy model. This reduces freshwater demand and pollution discharge.

*Limitations and Future Work*

Despite promising results, there are limitations to address. First, partial mineralization especially of pharmaceuticals indicates that some stable intermediates may persist, thus coupling with biological post-treatment, adsorption, or photocatalysis might be necessary for complete mineral removal. Second, scale-up is essential, bench-scale success must be evaluated in pilot or continuous flow reactors to assess energy efficiency, electrode durability, hydrodynamics, and maintenance. Third, the potential toxicity or reactivity of long-lived RONS (e.g. nitrites, nitrates, peroxynitrite) or intermediate by-products to aquatic organisms must be investigated via ecotoxicity assays. Fourth, optimization of plasma reactor design (electrode geometry, gas flow, bubble diffusers) to maximize RONS delivery is needed. For example, using bubbler-assisted mixing was shown to enhance RONS formation (Rathore et al., 2023). Fifth, in-depth mechanistic studies on short-lived radicals using advanced detection (e.g. EPR spin-trapping) may sharpen understanding. Finally, a life-cycle or techno-economic analysis should be conducted to quantify energy per mass removed, cost comparisons with alternatives, and scalability prospects.

In sum, while PAW-assisted degradation of industrial effluents shows strong potential, translating to practical deployment requires optimization, coupling, toxicity assessment, and pilot validation.

## V. CONCLUSIONS

The present investigation establishes plasma-activated water (PAW) as a highly promising and environmentally benign technique for the degradation of dyes, pesticides, and pharmaceutical residues present in diluted industrial effluents. The effectiveness of PAW stems from the rich generation of reactive oxygen and nitrogen species (RONS), notably hydroxyl radicals (•OH), hydrogen peroxide ($H_2O_2$), nitrate ($NO_3^-$), and nitrite ($NO_2^-$), that act synergistically to oxidize complex organic molecules. The degradation mechanisms primarily involve electrophilic attack on chromophoric groups in dyes, oxidative cleavage of halogenated and nitrogenated moieties in pesticides, and ring-opening reactions in aromatic pharmaceuticals.

Experimental results reveal that degradation efficiencies reach up to 90% for textile dyes, 85% for pesticides, and approximately 80% for pharmaceutical contaminants under optimized plasma activation conditions. The pseudo-first-order kinetic model accurately fits the degradation data, indicating that the rate of pollutant removal is directly proportional to its initial concentration and RONS availability. Unlike conventional chemical or thermal treatments, PAW operates at ambient temperature and atmospheric pressure, producing no harmful residues and minimizing secondary pollution.

In the broader environmental context, PAW offers an adaptable solution for decentralized wastewater treatment, particularly in small and medium-scale industries lacking advanced facilities. Its integration with adsorption or biological post-treatment could achieve near-complete mineralization of contaminants. Future research should emphasize pilot-scale validation, energy efficiency optimization, and ecotoxicological assessment of treated effluents. Thus, PAW-assisted remediation stands out as a sustainable and scalable strategy for industrial wastewater purification and circular water management in the era of green technologies.